\documentclass[11pt,twoside]{article}  
\usepackage{apn3conf}

\begin{document}   

%
%
%
%

\title{A new sample of young Planetary Nebulae }

%
%
%
\author{Grazia Umana, Luciano Cerrigone\altaffilmark{1},
			Corrado Trigilio} 
          \affil{Istituto di Radioastronomia CNR, 
                 Sezione di Noto, Italy}
		\author{R. Aldo Zappal\`a}
	\affil{Istituto di Fisica e Astronomia, Universit\`a di Catania, Italy}
          \altaffiltext{1}{Istituto di Fisica e Astronomia, Universit\`a di Catania, Italy}


%
%
\contact{Grazia Umana}
\email{g.umana@ira.cnr.it}

%
%
%
%
%

\paindex{Umana, G.}
\aindex{Cerrigone, L. }     
\aindex{ Trigilio, C.}
%
%

\authormark{ Umana, Cerrigone, Trigilio and Zappal\`a}

%
%



\begin{abstract}          
Despite of numerous efforts, the stage from Asymptotic Giant Branch (AGB)
to Planetary Nebulae (PN) is a poorly understood phase of stellar evolution.
We have therefore carried out interferometric (VLA) observations of a
sample of hot post-AGB stars, selected on the basis of their optical and infrared properties.
Ten sources, out of the 16 been observed, were detected. This indicates that most of our targets
are surrounded by  a nebula where the ionization has already started.
This definitively determines the evolutionary status of the selected sources and  provides
us with  a unique sample of very young Planetary Nebulae (yPNs).
\end{abstract}

%
%

\section{Introduction}
During the last few years many observational programs have been devoted to recognize new planetary and proto-planetary nebulae.
Such studies were aimed to understand the process of formation of PNs  by
discovering  new transition objects in the very  short phase between the end
of the AGB  and the onset of the ionization.

A small sub-group of B-type stars, called BQ[] stars, defined as
$B_\mathrm{e}$ with forbidden emission lines,
were recognized as potential candidates to be new transition objects
(Parthasarathy \& Pottasch, 1989) on the basis of their
IR excess. BQ[] stars, however, are not a well defined group, and there is still a controversy on their evolutionary stage.

\section{Observations and Results}

In order to find new very young Planetary Nebulae we have selected  a sample
of 16 hot post-AGB stars (BQ[]) from the most recent compilations, namely
Parthasarathy, (1993); Conlon et al., (1993) and  Parthasarathy et al.,
(2000).
All the selected  candidates  have high galactic latitude,
infrared excess, spectral type B1 I-II and the presence of  nebular emission lines in their
spectrum.
In particular the last two requirements maximize the possibility to detect a radio nebula.
The detection of radio nebulae associated to the selected targets
would definitively  assess  the evolutionary status of this kind of objects, producing
a unique sample of very young PNs.\\

The observations were carried out  by using the VLA in two different runs,  on June 8 and 10,  2001, at 8.4 GHz in CnB configuration.\\
We detected a total of 10 sources over 16 been observed.
Our results are summarized in Table~1, where the radio flux density, with
associated error ($\sigma$) and the rms of the map, are reported.
The error associated to the flux density estimation ($S$) is a combination of
the rms in the map plus the amplitude  calibration $error, \sigma_{cal}$,  
$\sigma = \sqrt{ (rms)^{2} + (\sigma_{cal} S)^{2}}$.

\begin{center}
\begin{table*}
\caption[]{Radio proprierties of observed stars}
~\\
~\\
\begin{tabular}{|ll|ccc|} \hline\hline
{\em Name} & IRAS   &   {\em Flux density} & {\em $\sigma$} & {\em rms}\\
           &        &    {\em mJy}         &    {\em mJy}   & {\em mJy}\\
\hline
           &        &                      &                &  \\
OY~Gem &  IRAS~06556+1623    &   0.55 & 0.03 & 0.02\\
 Hen 3-1347      &  IRAS~17074-1845        &   & & 0.03 \\
       &  IRAS~17203-1534      &  & 0.03\\
 LS 4331      &  IRAS~17381-1616    & 1.42  &  0.05 & 0.03\\
Hen 3-1475        &  IRAS~17423-1755    & 0.26  &  0.03 & 0.03 \\
SAO 209306        &  IRAS~17460-3114    & 1.29  & 0.05 & 0.04\\
V886 Her  & IRAS~18062+2410  & 1.46  &  0.05 &0.03 \\
 LS 63       &  IRAS~18371-3159    & 0.62  & 0.03 & 0.03 \\
 LS 5112      &  IRAS~18379-1707   &  &  & 0.03 \\
       &  IRAS~18435-0052    & & &  0.03\\
BD -11 4747        &  IRAS~18442-1144  & 19.21 & 0.60 & 0.05 \\
LS IV -02 29     &  IRAS~19157-0247    &       &      &  0.03\\
      		 &  IRAS~19336-0400    & 9.74  & 0.29 & 0.03\\
 LS II +23 17    &  IRAS~19399+2312    & & &0.03\\
 LS IV -12 111   &  IRAS~19590-1249    & 2.76  & 0.08 & 0.03 \\
LS II +34 26     &  IRAS~20462+3416    & 0.42  &  0.03 & 0.03\\

\hline \hline
 \end{tabular}
\end{table*}
\end{center}
\section{Discussion and Conclusions}
 In order to calculate distances to all the investigated
sources, we have chosen to apply, among different methods, that developed by
Tajitsu \& Tamura (1998) which  takes advantage of one characteristic
common to all the sources in our sample, namely the presence of a
strong  far-infrared flux as detected by IRAS satellite and gives as by-product the total far-IR flux ($F_{IR}$).\\
No systematic difference in distances between detected and non-detected sources have been found.
Therefore the non-detections are not related to larger distances but should be related to intrinsic characteristic of the source, such as the evolution of the ionization structure.

In order to check if our sample consists of young PNs we calculate some physical quantities, whose
values can help in understanding the evolutionary stage of the nebula.
Those are (Table 2): 
\begin{itemize}
\item the brightness temperature, that for young nebulae should be of the order of $T_{B}\sim 10^{3} K$
\item the Emission measure, that for young nebulae should be of the order of $EM \sim 10^{8} cm^{-6} pc$
\item the Infrared Excess, that for young nebulae should be $IRE \geq 1$
\end{itemize}
\begin{table*}
\caption{Summary of nebular characteristics of detected targets.
IRE derived following  Pottasch (1984);
The mean emission measure (EM) has been calculated from
the formula of Terzian \& Dickey (1973)}
~\\
~\\
\begin{tabular}{|lcccc|}
\hline \hline
 IRAS ID & IRE & Diameter & $T_\mathrm{B}$ &  $EM$  \\
 &   & [arcsec] & [K] & [$10^{4}\mathrm{cm}^{-6}$pc] \\
\hline
\emph{06556+1623} & 194 & 2.1         & 2.3        & 6.3  \\
\emph{17381-1616} & 31  & $\leq 2.0$  & $\geq 8.9$ & $\geq 18.8$  \\
\emph{17423-1755} & 2984 & $\leq 2.0$  & $\geq 1.6$ & $\geq 3.4$  \\
\emph{17460-3114} & 248  & 1.1         & 27         & 56.6 \\
\emph{18062+2410} & 106  &  $\leq 2.0$ & $\geq 9.1$ & $\geq 19.3$ \\
\emph{18371-3159} & 187 &  $\leq 2.0$ & $\geq 3.9$ & $\geq 8.2$  \\
\emph{18442-1144} & 11  & 1.8         & 148        & 314.8 \\
\emph{19336-0400} & 14   &  1.5        & 108        & 229.8 \\
\emph{19590-1249} & 21   & 1.9         & 19.1       & 40.6 \\
\emph{20462+3416} & 186  &  2.2        & 2.2        & 4.6 \\
\hline\hline
\end{tabular}\end{table*}
Our results have been compared to those obtained by Aaquist and Kwok (1991, AL91), who observed with the VLA at 15~GHz a sample of yPNs.\\
 For  the AK91 sample $T_{B}$ and $EM$ are  systematically higher than those of our sample;
this could imply that our sample consist of more evolved PNs.
On the contrary, IREs for our sample are systematically higher than those reported by AK91, implying that our sample 
is formed by PNs particularly young.\\
  This apparent contradiction is further complicated by the fact the infrared properties of both samples are quite similar: sources belonging to differenr samples occupy the same region in the IRAS color-color diagram and dust temperatures of both samples have quite similar distributions, implying analogous dust characteristics. 

A possible cause of lower $T_{B}$ and $EM$ of our sample can be a systematic effect due to the different spatial resolution used in the two surveys, as both $T_{B}$ and $EM$ are function of the source angular size ($ \propto \theta^{-2}$). However, only 4 out of 10 detected sources were not resolved.

The apparent contradiction can be explained if we assume that sources of both samples are in the ionization bounded phase of radio nebula evolution, but our sample is less evolved and is characterized, on the average, by a lower radio luminosity when compared to the AK91 sample.
Consistently, ionized masses for our sample are in the range $3 \times 10^{-5}-1.6  \times 10^{-3} M_{\odot}$, much lower than typical values for evolved nebulae (Pottasch, 1984).

The detection of free-free radio emission in 10 of the observed sources indicates that ionization is already started in their circumstellar shells.  The detected sources are in the very early stage of PNs evolution and constitute a unique sample to be studied to shed light on this quite poorly understood phase of stellar evolution.

Successive multi-frequency  and high-resolution radio observations will allow to fully characterize the radio properties of these new objects.

\acknowledgments
The Very Large Array is a facility of the
National Radio Astronomy Observatory which is operated by Associated Universities, Inc. under cooperative agreement with the
National Science Foundation
%
%
%
%


\end{document}